\documentclass[conference]{IEEEtran}

\usepackage[utf8]{inputenc}
\usepackage{listings}
\usepackage{xcolor}
\usepackage{graphicx}
\usepackage{svg}
\usepackage{tikz}
\usepackage{subcaption}
\usepackage{pifont}

\usepackage{amssymb}
\usepackage{tikz}
\usepackage{pgf-umlsd}
\usetikzlibrary{automata, positioning, arrows}
\usepackage{url}

\tikzset{
->,
>=stealth,
node distance=3cm,
every state/.style={thick, fill=gray!10},
initial text=$ $, 
}

\title{Reducing Usefulness of Stolen Credentials in SSO Contexts}
\author{
    \IEEEauthorblockN{Sam Hays, Michael Sandborn, Jules White}
    \IEEEauthorblockA{
        \textit{Department of Computer Science, Vanderbilt University, Nashville, TN, USA} \\
        \{george.s.hays, michael.sandborn, jules.white\}@vanderbilt.edu}
}

\begin{document}

\maketitle
\pagestyle{plain}

\section{Abstract}
\label{Abstract}
Approximately 61\% of cyber attacks involve adversaries in possession of valid credentials. Attackers acquire credentials through various means, including phishing, dark web data drops, password reuse, etc. Multi-factor authentication (MFA) helps to thwart attacks that use valid credentials, but attackers still commonly breach systems by tricking users into accepting MFA step up requests through techniques, such as ``MFA Bombing'', where multiple requests are sent to a user until they accept one. Currently, there are several solutions to this problem, each with varying levels of security and increasing invasiveness on user devices.  This paper proposes a token-based enrollment architecture that is less invasive to user devices than mobile device management, but still offers strong protection against use of stolen credentials and MFA attacks.

\section{Introduction}
\label{Motivation}
Single Sign-on (SSO) is an identity management approach adopted by modern businesses to enable authentication to multiple services through a single set of credentials. The principal idea is allowing a company to control the login process and associated authentication requirements for its users on behalf of third-party services for which the company could otherwise exert no control. Today, SSO typically refers to the login workflow associated with corporate users logging into third party applications that are not controlled by the corporate information technology staff, usually with a web browser\footnote{We differentiate this from \emph{Social Logins} such as ``Sign-In with Gmail'' and ``Sign-In with LinkedIn'' which do provide an SSO-Experience but falls in the B2C space rather than the B2B space.}. To support this model, a number of protocols exist under which Single Sign-On~\cite{ssosrv} may be implemented, such as SAML 2.0~\cite{saml2}, OpenID Connect~\cite{openID}, and even protocols unrelated to web technologies, such as Kerberos~\cite{Kerberos}.

When login is allowed to happen from anywhere in the world, as opposed to being restricted to from a specific network, it also increases the attack surface and threat model complexity. A globally-accessible login service has greatly exacerbated the problem of Phishing because users had previously needed to be on the physical network of the enterprise campus to leverage their stolen credentials; now the campus has grown to the entire planet. Indeed, Alkahalil et al. ~\cite{phin} has shown that Phishing attacks to obtain credentials are getting to the point of sophistication that they are indistinguishable from real communications from business partners. Furthermore, Thomas et al. ~\cite{brphma}, in their analysis of the data from March 2016-2017 including 1.9 billion username/password pairs from data breaches, shows that  bad actors within the analysis were collecting \textbf{nearly 235K potentially valid credentials every week}.

To address this global exposure, companies often require a second form of authentication such as mobile authenticator/MFA whereby a user must possess an object, such as a mobile phone, which has been associated to the user and presents a one-time key or prompt during the login process. 
These additional controls can be effective in reducing successful usage of a compromised account by an adversary, although it is by no means guaranteed as seen from the very public breaches involving so-called ``MFA Bombing'' ~\cite{2fabomb}.

This paper offers an architecture to limit the exposure of a login endpoint to devices that have been previously vetted by a well-defined corporate security policy and corresponding enrollment endpoint, such that an attacker who is in possession of corporate credentials cannot leverage them to sign into any service protected by SSO. 

The primary insights of this paper can be enumerated as follows:

\begin{enumerate}
    \item The entire user population behaving in a security-compliant way is extremely unlikely.
    \item Multi-factor authentication often fails if a user does not behave in accordance with security policy.
    \item We may add an additional security challenge to adversaries with minimal user impact (other than a per-device enrollment step) such that the adversary cannot access the sign-on portal to use credentials that they may have obtained.
    \item The additional challenge does not adversely impact the usability of daily user authentication activities, while protecting against inadvertent MFA step-ups.
\end{enumerate}

With these tenets in mind, we propose the following solution.

\begin{itemize}
    \item A (thoroughly secured) lightweight one-time per-device enrollment endpoint to register a device, resulting in a signed JSON Web Token (JWT)~\cite{jwt} which shall be verified to make available the login page.
    \item A route to serve the SSO credential login page only in the case that the signed JWT provided by the requesting browser is both present and valid.
    \item A flexible and extensible implementation to introduce granular control of security policies and authorization artifacts for addressing ``Semi-Untrusted Users.''
\end{itemize}

The contributions of this work are the following: 

\begin{itemize}
    \item We characterize the \emph{Semi-Untrusted User problem} which refers to users who may act in a way inconsistent with corporate security policies (e.g., leak credentials via phishing, mistakenly approve an MFA login request resulting from a malicious actor providing stolen account credentials, etc.)
    \item We define an architecture for addressing this problem that is lightweight, flexible, and extensible and reduces the usefulness of stolen credentials to an attacker.
    \item We ensure the TULIP methodology is not strongly bound to any single protocol (e.g., SAML2/OIDC) but may augment the strength of any.
\end{itemize}

The remainder of this paper is organized as follows:
Section~\ref{Prior Work} compares our approach to existing approaches.
Section~\ref{TULIP} outlines our solution approach, which we call ``TULIP'' (The Unavailable Log-In Page).
Section~\ref{eval} provides an analysis and evaluation of the security characteristics of TULIP.
Section~\ref{conclusion} Provides concluding remarks.

\section{Existing Techniques and Prior Work}
\label{Prior Work}

This section outlines the related work of device trust as a means to reduce successful compromise of corporate accounts in cases where credentials are in the possession of an attacker. Establishing a device as being ``trusted'' is achievable through several means, each with their own merits and challenges. Our method involves a strongly protected enrollment endpoint to grant tokens that must be present for the login screen to be rendered and process login requests. Here we consider existing device trust mechanisms and compare them to our proposed solution.

\subsection{Mobile Device Management (MDM)}

        Mobile Device Management~\cite{mdmsec}, \cite{cloudmdm}, \cite{mdmrequirem}, \cite{mdmTech} is a process by which software is loaded onto a device and can then interrogate, control, and otherwise influence the device's behavior depending on the software and associated policies.
        
        MDM is perhaps the most intrusive form of establishing trust with a device and is typically targeted at Smart Phones and Tablets. This solution is often seen as reasonable in cases where a company provides the devices to the end users, but in circumstances where the workforce supplies their own equipment via a ``Bring Your Own Device'' (BYOD) policy, the requirement to install corporate software that can control, modify, and even ``kill'' the device may be seen as corporate overreach. 
        
        As a consequence of the potential blowback of instituting such a policy, many companies elect to forgo this route. In these cases, another less intrusive method is employed, or there is no enrollment process and other data analysis is performed to detect-and-respond to potential issues rather than preventing them from the outset.

        Our approach relies on a lightweight enrollment protocol to satisfy the trust of a device, albeit to a lesser degree than MDM.

\subsection{X509 Certificates}

        X.509 certificates~\cite{509risk}, \cite{pkcmgmt}, \cite{certMemo} are cryptographic certificates based on the Public Key Infrastructure (PKI) standards. These certificates can be loaded onto a device and later used to authenticate the device. X.509 have excellent security properties and are often isolated from other system resources \cite{enclav}, making their theft extremely difficult. However, three core challenges make X.509 usage for SSO relatively rare in industry: infrastructure management expertise, certificate maintenance and hygiene, and end-user training.
        
        First, installation, configuration, and maintenance of a PKI can often be outside the expertise of small to medium sized companies. Larger companies often use external certificate authorities where a cost is involved in each certificate procured, making this solution potentially cost-prohibitive. 
        
        The second problem is certificate maintenance and hygiene. Expertise must be maintained for every supported device type, and a certain risk tolerance for claims of malfunctioning devices being a result of such installations. Once certificates are loaded, then a certificate revocation list \cite{certMemo} must be maintained. Finally, a policy must be established to revoke certificates when devices are decommissioned (e.g., lost, stolen), and when a user is no longer associated with the company. 
        
        The third issue represents somewhat of a regression to the original problem we aim to address: end-user training. The web-browser experience of using X.509 certificates ranges from somewhat non-obvious to completely broken, depending on the device OS, browser, etc. Different devices may display the prompts requesting a user to select which certificate to present to the SSO page differently (or not at all). Users must be trained to handle these prompts appropriately and this kind of training can be expensive, difficult, and often ineffective as well as onerous for staff.

\subsection{Device Profiling}

 Device Profiling is another common technique used by companies hesitant to employ stricter controls on the user population. In this paradigm, it is common for third-party services to use JavaScript in the browser to attempt to profile the device and ascertain the probability that it is a known device and whether or not it is associated with a specific user. In some cases, a cookie may be set so a ``verified'' user can be identified easily on subsequent sign-on attempts. Typically, after the profiling step, a decision about whether to allow the sign-on action to continue, to force a second factor, or to take some other action can be made.
    
    This approach can be attractive for companies who have achieved a level of sophistication (and budget) for the introduction of this kind of tooling in the environment. Nevertheless, the profiling does not stop the user-behavior problem described in this work and indeed, the authors have observed these kinds of solutions implemented that still fail when a user effectively acts as an adversary by completing the second factor blindly.
    
    Because our proposed solution makes the sign-on screen effectively invisible except to enrolled devices, it may do a better job of reducing risk. Additionally, the profiling method and our solution can be used in tandem as a defense-in-depth strategy.

\section{TULIP}
\label{TULIP}

    The TULIP architecture defines a mechanism by which the SSO Login Page will not be available to any device which has not performed the enrollment actions. As a consequence, any actor in possession of credentials will be unable to use them for login purposes without a token granted during the enrollment process. The specifics of this protocol are covered in detail in this section.
    
    \subsection{Threat model and assumptions}
        We assume an attacker is already in possession of the user credentials of the victim at the outside of our analysis. We assume the user may or may not have MFA configured on their account. If they do, TULIP will mitigate MFA-Bombing style attacks indirectly by preventing the user from attempting a (successful) login in the first place.

    \subsection{The Enrollment Endpoint}
    
        The linchpin of this solution is the enrollment endpoint. The differentiating characteristic of this endpoint versus the standard sign-on endpoint is the exchange of ease-of-use with more onerous security requirements. More concretely: To achieve a positive user experience and welcome adoption of the standard SSO workflow, a balance must be reached between security and usability. This balance can be skewed towards security during the enrollment process because it should be a one-time (or very infrequent) process per device, comparable to configuring WiFi for a corporate network.

The enrollment endpoint will, upon successful execution of the process, yield a JWT to be stored in the user's browser. We recommend using a persistent cookie with adequate security mechanisms enabled (e.g., \texttt{httponly, secure}, and \texttt{samesite} set to \texttt{lax}). This JWT will be digitally signed and verified by the ``true SSO endpoint'' (i.e., the one configured and used by Service Providers) prior to rendering the sign-on page. In the case that a request arrives at the server with a missing or invalid JWT, the SSO Login Page will not be rendered nor available to the party attempting the login request. In the event the client manipulates a request to \texttt{POST} the form through some other means, such as cURL, the JWT would need to be both present and valid before any login processing with form data will occur on the server side. Therefore, with the absence of a token, this control ensures the login page is unavailable using the ``front-door'' (i.e., web browser) and any authentication attempts using secondary mechanisms will also fail to process on the back-end.

Our approach introduces the requirement that a device attempting to sign-in to the system must have completed the enrollment process beforehand. Consequently the enrollment endpoint must be more stringently protected than the standard login endpoint. The level of protection of this endpoint is left to the analysis of the implementer but recommendations follow as a suggested set of starting point considerations.

\begin{enumerate}
            \item Ensure the enrollment endpoint is only available within the corporate network. In this case, a person must physically bring their device to network (e.g., corporate office) and complete the enrollment from there. This restricts the enrollment capabilities from geospatial (i.e., the endpoint is not on the Internet for the world to access) as well as authorization (i.e., a user must have access to the building as well as access to the corporate network) perspectives.
            \item An addition to the above requirement could be allowing enrollment over VPN. This still makes invalid enrollments very difficult, especially if the VPN uses certificates + TPMs and other strongly secured authentication and authorization mechanisms.
            \item Require the user to be a member of a specific group in the technology backing the enrollment. For example the user may have to call a help-desk, verify themselves, and then be temporarily added to a group during the enrollment period and then removed from the group upon successful enrollment. Only members of this group can achieve enrollment. 
            \item Prompt the user for secondary information beyond credentials (e.g., employee ID).
            \item Require an MFA during enrollment. Ensure the user must unlock their MFA device and manually enter the (T)OTP to achieve enrollment.
        \end{enumerate}

        \begin{figure}[ht]
            \centering
            \begin{tikzpicture}
                \node[state, initial] (qr) {$q_r$};             
                \node[state, below right of=qr] (qt) {$q_\tau$};
                \node[state, accepting, right of=qt] (qd) {$q_\delta$};
                \node[state, above right of=qr] (qe) {$q_\epsilon$};
                \node[state, accepting, right of=qe] (qa) {$q_\alpha$};
                \draw
                    (qr) edge[above] node{$\varnothing$} (qe)
                    (qr) edge[above] node{$\tau$} (qt)
                    (qt) edge[left] node{invalid} (qe)
                    (qt) edge[bend left, right] node{valid} (qa)
                    (qe) edge[bend left, right] node{decline} (qd)
                    (qe) edge[above] node{accept} (qa)
                    ;
            \end{tikzpicture}
            \caption{Outcomes of the Enrollment Procedure}
            \label{fig:enrollment}
        \end{figure}
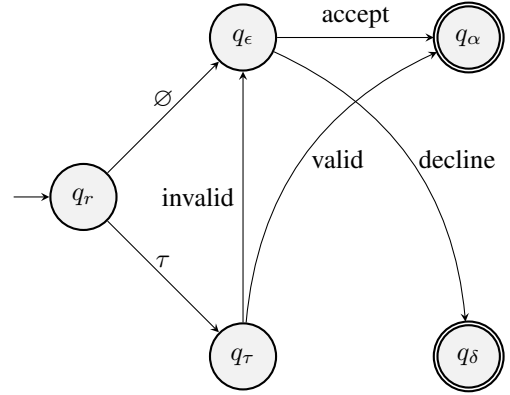

\begin{figure}
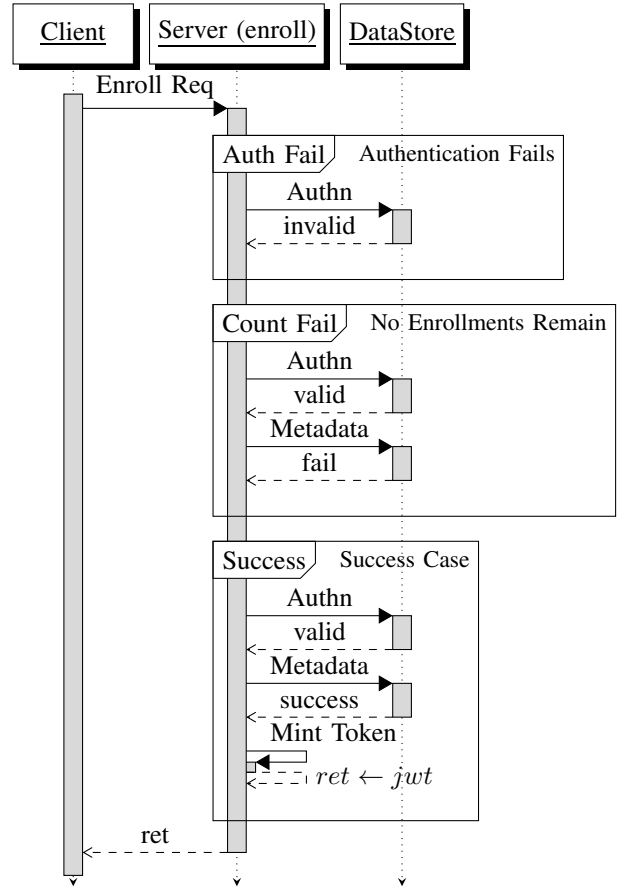

    \centering
    \begin{sequencediagram}
        \renewcommand\unitfactor{0.45}
        \newthread{client}{Client}
        \newinst{server}{Server (enroll)}
        \newinst{data}{DataStore}
        \begin{call}{client}{Enroll Req}{server}{ret}
            \begin{sdblock}{Auth Fail}{Authentication Fails}
                \begin{call}{server}{Authn}{data}{invalid}\end{call}
            \end{sdblock}
        
            \begin{sdblock}{Count Fail}{No Enrollments Remain}
                \begin{call}{server}{Authn}{data}{valid}\end{call}
                \begin{call}{server}{Metadata}{data}{fail}\end{call}
            \end{sdblock}
    
            \begin{sdblock}{Success}{Success Case}
                \begin{call}{server}{Authn}{data}{valid}\end{call}
                \begin{call}{server}{Metadata}{data}{success}\end{call}
                \begin{call}{server}{Mint Token}{server}{$ret \leftarrow jwt$}\end{call}
            \end{sdblock}
            
        \end{call}
    \end{sequencediagram}
    \caption{Enrollment Flow}
    \label{fig:seq_enrollment}
\end{figure}

    In Figure \ref{fig:enrollment} we understand that $q_r$ represents the initial request into the Enrollment Endpoint. If a token is not present ($\varnothing$) then the system transitions to $q_\epsilon$ for the enrollment procedure (represented by Sequence Figure~\ref{fig:seq_enrollment}). Upon successful authentication the system issues its token at state $q_\alpha$. If the authentication procedure fails then the state ends at $q_\delta$. Similarly if at $q_r$ the a token ($\tau$) is already present then it is checked for validity. If it is already valid then we do not re-issue a token, and instead transition to a completed state at $q_\alpha$, otherwise the token is discarded and state transitions to $q_\epsilon$ for the enrollment process (which is convenient in cases where token revocation has occurred).

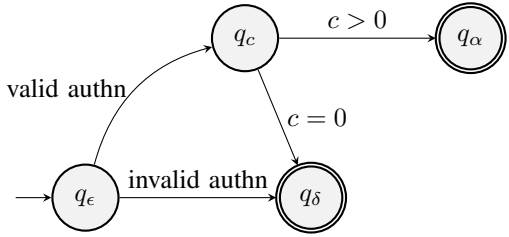
\begin{figure}
            \centering
            \begin{tikzpicture}
                \node[state, initial] (qt) {$q_\epsilon$};             
                \node[state, above right of=qt] (qc) {$q_c$};
                \node[state, accepting, right of=qt] (qd) {$q_\delta$};
                \node[state, accepting, right of=qc] (qa) {$q_\alpha$};
                \draw
                    (qt) edge[bend left, left] node{valid authn} (qc)
                    (qt) edge[above] node{invalid authn} (qd)
                    (qc) edge[above] node{$c > 0$} (qa)
                    (qc) edge[right] node{$c = 0$} (qd)
                    ;
            \end{tikzpicture}
            \caption{Enrollment Validation}
            \label{fig:token_validation}
        \end{figure}

    The state $q_\epsilon$ bears further discussion. Within this state (Figure \ref{fig:token_validation}), the enrollment is validated \& protected in two ways. First, the user in prompted for authentication. If this fails, then we shift to $q_\delta$. If the user successfully authenticates themselves, then the protocol will check to ensure the user has not consumed all available enrollments as described in ~\ref{token_limits}. If they have, then we still transition to $q_\delta$. We note that TULIP allows for arbitrary challenges in the enrollment process to accommodate any security requirements.
    
  Given the importance of the enrollment endpoint in the prevention of subsequent unauthorized logins, these recommendations are intended to inspire deep consideration of the security requirements for implementers of the protocol. TULIP does not prescribe the implementation details, so it is the responsibility of the implementer to design procedures and policies that satisfy the security requirements of their environment. We note that TULIP, even if poorly implemented, cannot reduce the SSO security posture if the SSO page is already publicly available and subject to attackers and fraudsters. Nevertheless, for effective reduction of the successful use of compromised credentials, a strong enrollment endpoint must be designed and implemented.

        The following scenario illustrates these ideas.
        \begin{itemize}
            \item A user enrolls their device via the enrollment endpoint \\(\texttt{https://sso.company.com/enroll})
            \item The newly minted JWT is stored as a cookie
            \item Upon SP-Initiated or IdP-Initiated login, the browser lands on \texttt{https://sso.company.com} and presents the cookie in the HTTP request
            \item If the JWT is valid, then the login page is rendered (otherwise it is not)
            \item When a JWT is presented in the request, it is checked for validity and revocation criteria prior to the login page being made available or any data from the client being processed
        \end{itemize}

    \subsection{The Login Endpoint}
    \label{login_endpoint}

\begin{figure}[ht]
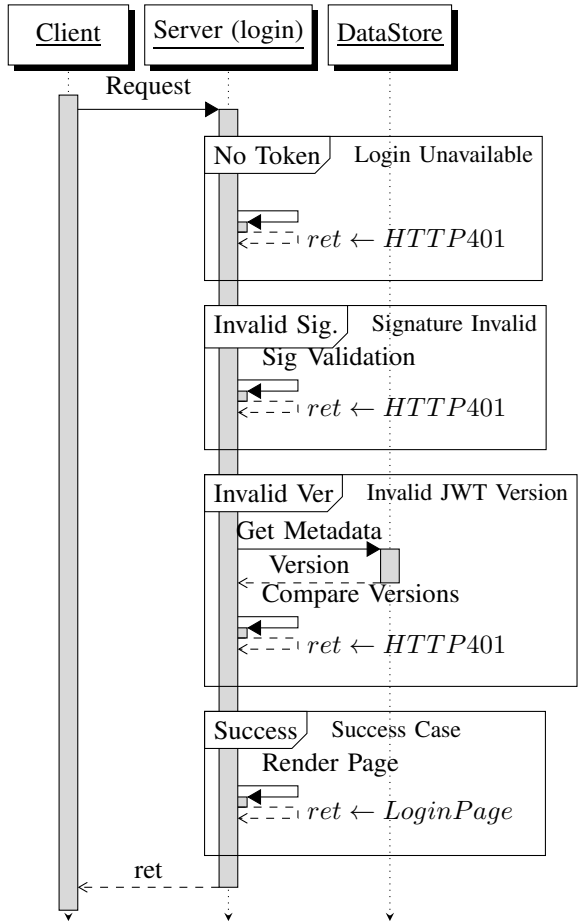

    \centering
    \begin{sequencediagram}
        \renewcommand\unitfactor{0.45}
        \newthread{client}{Client}
        \newinst{server}{Server (login)}
        \newinst{data}{DataStore}
        \begin{call}{client}{Request}{server}{ret}
            \begin{sdblock}{No Token}{Login Unavailable}
                \begin{call}{server}{}{server}{$ret \leftarrow HTTP401$}\end{call}
            \end{sdblock}
        
            \begin{sdblock}{Invalid Sig.}{Signature Invalid}
                \begin{call}{server}{Sig Validation}{server}{$ret \leftarrow HTTP401$}\end{call}
            \end{sdblock}

            \begin{sdblock}{Invalid Ver}{Invalid JWT Version}
                \begin{call}{server}{Get Metadata}{data}{Version}\end{call}
                \begin{call}{server}{Compare Versions}{server}{$ret \leftarrow HTTP401$}\end{call}
            \end{sdblock}

            \begin{sdblock}{Success}{Success Case}
                \begin{call}{server}{Render Page}{server}{$ret \leftarrow Login Page$}\end{call}
            \end{sdblock}
            
        \end{call}
    \end{sequencediagram}
    \caption{Login Flow}
    \label{fig:seq_login}
\end{figure}

    The Login Endpoint is the central focus of protection of the TULIP protocol. The objective is to make this page unusable to an attacker in possession of credentials since they should not have an enrolled device, and the protocol limits the enrollment procedure by both having stricter enrollment requirements, and capping the amount of devices any individual user may enroll (described in ~\ref{token_limits}). We see in Figure ~\ref{fig:seq_login} that only in the condition that the token is present (device has been enrolled), has a valid signature (the token is unmolested), a correct version (the token has not been revoked), does the Login Page render and make available the form to which a user may attempt the login process. We note that all of these protections exist prior to the login process itself and do not provide a login mechanism themselves, giving the attacker no advantage.

\begin{figure}
            \centering
            \begin{tikzpicture}
                \node[state, initial] (qr) {$q_r$};             
                \node[state, above right of=qr] (qv) {$q_\upsilon$};
                \node[state, accepting, right of=qr] (qd) {$q_\delta$};
                \node[state, accepting, right of=qv] (qa) {$q_\alpha$};
                \draw
                    (qr) edge[above] node{$\tau$} (qv)
                    (qr) edge[above] node{$\varnothing$} (qd)
                    (qv) edge[above] node{valid} (qa)
                    (qv) edge[right] node{invalid} (qd)
                    ;
            \end{tikzpicture}
            \caption{Token Validation \& Login Page Availability}
            \label{fig:token_enrollment_check}
        \end{figure}
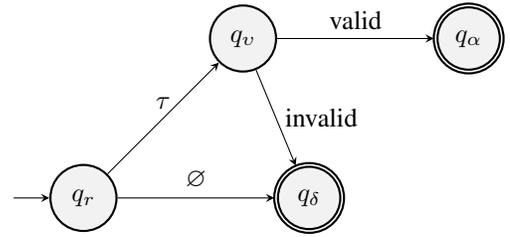

    We may also visualize this as in Figure ~\ref{fig:token_enrollment_check} the state of the system changes to $q_\upsilon$ if a token is valid. It is this state in which all token checks described here are undertaken to decide what the appropriate response to the user might be (e.g., returning the login page or an HTTP 401).
    
    \subsection{Token Revocation}
    \label{trevo}

        Because the tokens are persistent and stored in the browser, a mechanism must be implemented to allow for token revocation if evidence of a breach has been identified~\cite{memsec}.
        
        To accomplish this, the authors recommend a counter variable stored as a user-attribute in the backing identity source of record (e.g., LDAP, Database, Active Directory, etc). We call this field the \texttt{sso-jwt-version}. When a user enrolls, the \texttt{sso-jwt-version} field's value is set in the JWT. 
        
        For example, suppose a user object's \texttt{sso-jwt-version} is configured as ``2''. When the JWT is minted for enrollment, a \texttt{version} field is included with this value. When the user begins the login process and presents the JWT cookie (prior to rendering the login screen), the version in the JWT will be compared against the version in the backing directory service and fail if they do not match\footnote{Astute readers may notice that we are somehow identifying the user via the JWT prior to authentication. This is explained in the next section.}.
    
        In this way, any JWT with a version not matching the \linebreak \texttt{sso-jwt-version} back-end attribute will be considered revoked. Thus, all previously issued tokens can be revoked by simply incrementing or otherwise changing this value.
        
        The implementer would also need to establish a revocation procedure if evidence suggests a bad actor has obtained a token. One such procedure might involve help-desk or system administrator scripts which simply increment the \texttt{sso-jwt-version}, rendering useless all previous tokens. It may also be appropriate to use a global revocation action after a prescribed period of time if the company wishes to occasionally require global re-enrollment to clear stale devices.

         \subsection{Identifying Users}
        It is convenient for the SSO system to ascertain the identity of a user by metadata contained in the JWT for processing. However, we do not wish to include any sensitive information (including such things as username) within the token itself. To address this concern, we introduce the attribute called \texttt{sso-jwt-ouid}.
        
        The \texttt{sso-jwt-ouid} (``opaque user identifier'') is a user attribute in which a unique pseudo-random value (e.g., UUID or GUID) is stored on the backing data store. This value is also written to the JWT when it is minted so that the JWT now includes \texttt{version} and \texttt{ouid} values. With this in place, the token verification can perform the following actions prior to rendering a login screen:
        
        \begin{enumerate}
            \item Check for the presence of a JWT (requires no back-end data store calls)
            \item Validate the signature of the JWT (no back-end calls)
            \item Validate the version of the token (one back-end call)
        \end{enumerate}
    
        The mechanism used for step 3 works in the following way. When the token arrives and has been validated (steps 1 and 2), the SSO environment reads the \texttt{ouid} and \texttt{version} attributes. The back-end uses the ouid to find the actual user object (e.g., assuming the SSO web server read in the JWT's ouid value as ``ouid'', then an LDAP query such as \texttt{(sso-jwt-ouid=\$ouid)} could be employed). Once the user object has been identified, then all relevant attributes, such as \texttt{sso-jwt-version} become available from the user object\footnote{The ouid lookup and collection of attributes can usually all occur with one back-end call to reduce network requests.}.
        
        Finally, the version attribute is checked to determine if the token has been revoked.
        
         \subsection{Token Limits}
         \label{token_limits}
        
        Typically, users do not have an infinite number of devices which should be eligible for SSO. Instead, they might have as few as one but more often almost fourteen ~\cite{taylor_2023} -- such as a phone, tablet, and laptops. Once these devices are enrolled, then no more devices should be able to obtain a token. 

        This allows for a number of flexible enrollment policies to be employed. For example, a company might ask each user to specify the number of enrollable devices they expect to use and configure the system to only allow that number. Another possibility is that the help-desk might only allow new enrollments with the opening of a service ticket.
        
        Whatever the mechanism used to limit devices, this is a security property that can prevent rogue enrollment for a token and further prevent identity abuse.
    
        To achieve this we introduce an attribute called \texttt{sso-jwt-count}. This value is set to some positive integer (e.g., 3) on a user's object. Any time a new device successfully goes through the enrollment process, the count is decremented. If it is at zero when a new enrollment is attempted, the enrollment must fail.
        
        If a user were to get a new device, then the help-desk or administrative staff could increment the count for the user to allow the new device to obtain its token (optionally also revoking the previous tokens if needed). The key insight is that when a user has consumed all of their allotted \texttt{sso-jwt-counts} and the value is at zero, no other device may be enrolled for that user.

\section{Evaluation}
\label{eval}

\begin{table*}[t]
\centering
\begin{tabular}{ |c|c|c|c|c| }
\hline
Attribute & SAML & SAML+MDM & TULIP (ours) & X.509 \\
\hline
 Does not require registration & \checkmark & - & - & -\\
 Is not globally accessible & - & \checkmark & \checkmark & - \\
 Is not limited by Device type/OS & \checkmark & - & \checkmark & \checkmark \\
 User is pre-verified & - & \checkmark & \checkmark & \checkmark \\ 
 Provides a trust revocation mechanism & - & \checkmark & \checkmark & \checkmark \\ 
 Does not require additional software & \checkmark & - & \checkmark & - \\
Provides side channel indicators of compromise & - & \checkmark & \checkmark & - \\
\hline
\end{tabular}
\caption{Comparison of TULIP's attributes to existing approaches for Semi-Untrusted User behaviors.}
\label{comparison}
\end{table*}

This section discusses how TULIP limits attacker capabilities in the context of an attack in which the attacker is in possession of the victim's credentials. We compare properties of TULIP to existing approaches, and introduce quantities for attacker success with and without the use of TULIP based on user behavior. 

Consider a company with $k$ employees who leverage an SSO login endpoint for cloud-base services. Some number $n$ ($n \ll k$) of these users have had their credentials (username/password pair) leaked via password reuse, phishing, or other means. As has been seen in public breaches from large technology companies, we know that an attacker repeatedly attempting to log in with these credentials will eventually cause a compromised user to accept the MFA request, allowing the attacker to gain access. In this scenario and given enough time, a breach will occur due to human error.

An implementation of TULIP, in the same scenario, does not result in a breach due to human error. There are still $k$ users of which $n$ have had their credentials compromised. However as part of the TULIP protocol, the $k$ users have enrolled their devices and are in possession of the token without which, the login cannot be attempted because the login page is not accessible. Without a login attempt, there cannot be a MFA notification. In this case, the $n$ compromised users would have to also have a second form of compromise (e.g., a stolen device + the thief having gained access / logged in, installed a RAT or Trojan Horse Program), any of which would have itself been a full compromise. We therefore assert that to breach TULIP would be tantamount to having physical access to the user's device, rather than only access to their credentials.

To the best of our knowledge, there is not a similar approach as TULIP in any existing approaches of which we are aware. Therefore, we compare our approach to technologies that are used in similar solutions to the Semi-Untrusted User problem. The comparison of our approach is summarized by Table~\ref{comparison}. To reiterate, TULIP ensures the worst case scenario is a regression to the original problem we aim to solve: to prevent successful usage of stolen credentials by an attacker.

\section{Conclusion}
\label{conclusion}

We introduce the ``Semi-Untrusted User Problem'' which refers to users who may behave in a way inconsistent with security policies and best practices. We discuss how such users may inadvertently allow attackers into a network even in the presence of second factors such as MFA due to errant behavior resulting from poor training, ``step-up fatigue'', and other factors contributing to poor judgement.

To combat this issue, we propose TULIP which stands for The Unavailable LogIn Page and solves the Semi-Untrusted User Problem by requiring a near-one-time user-enrolled device to request the login form to input credentials. Device enrollment is made possible using JSON Web Tokens (JWTs) and allows for the identity of enrolled devices to be asserted on login requests. The device enrollment step prevents an erroneously approved login request from completing a phishing attack since a malicious actor cannot initiate a login request without possessing an enrolled device.

We provide the enrollment and sign in flows for TULIP as well as the possible execution paths and how they are addressed with a generic failure response if a user attempting login fails to provide a valid token or an enrolled device. We also compare TULIP to existing approaches and develop expressions for estimating the efficacy of TULIP based on assumptions about user behavior and phishing prevalence.

We expect TULIP to increase both the cost to attackers to carry out successful phishing attacks as well as the security level of unsuspecting users by requiring the validation of a per-device persistent cookie to distinguish authorized login requests.

\subsection{Limitations}
    
    For users that are privacy-conscious to the extent of removing persistent cookies from their browser after a session (e.g. by clearing the browser cache), TULIP may be inconvenient since device enrollment is maintained with persistent cookies. In this case, we recommend the use of a separate browser for activities using the services provided by the SSO group.
    
    Some applications which use an embedded browser during authentication (e.g., Microsoft Outlook), may run the browser in an isolation mode whereby cookies are not shared, and thus the device registration JWT will not be available. These applications must be considered somewhat separately from the general SSO flows that are entirely browser-based. One possibility is for the IdP configuration to have a separate login screen for these connections with more onerous login requirements.
    
    If an attacker were to have a remote access tool (``RAT'') installed or could otherwise gain physical access to the computer, then the cookie could be viewed and copied from the browser's storage. This limitation is acceptable as it is generally understood that physical access (or virtually physical access in the case of RATs) is tantamount to a full breach with administrative privileges.

\subsection{Extensions}
    The TULIP protocol is intended to significantly increase the difficulty of leveraging stolen credentials by an attacker. This is accomplished by using an enrollment endpoint and storing the cookie in the browser in such a way that it cannot be stolen by common tactics (e.g., XSS). However, as browser technology changes and unique company needs are identified, this protocol allows for easy extensions. For example, a company may decide that they wish to add an attribute representing the User Agent of the browser during enrollment. They could write this attribute to the value and compare in all subsequent requests to ensure the browser is the same, etc.
    
    Similarly, there are emerging technologies such as WASM~\cite{wasm} to extend browser functionality. With these and other hardware technologies such as TPMs~\cite{tpms}, it becomes easier to store secrets in an even more secure manner. One such example is Google's TPM-JS~\cite{tpmjs}. This library, at the time of this writing, allows for browser-based access to TPMs via WASM. This type of technology innovation could, as TPMs become ubiquitous on common devices, further increase the security of the stored tokens by establishing trust at the hardware level.

\bibliographystyle{acm}
\bibliography{citations}
\section{Appendix 1: Figures}
\label{Apendix 1}

\end{document}